# Denial of Service Attack: Analysis of Network Traffic Anomaly Using Queing Theory

Neetu Singh, S.P. Ghrera and Pranay Chaudhuri

**Abstract**—Denial-of-service (DOS) attacks increasingly gained reputation over the past few years. As the Internet becomes more ubiquitous, the threat of the denial-of-service attacks becomes more realistic and important for individuals, businesses, governmental organizations, and even countries. There is intensive need to detect an attack in progress as soon as possible. The efficiency of diagnosing the DOS attack using concepts of queuing theory and performance parameter of the system has been investigated in the present work, as the servers definitely have some mechanisms to store and process the requests. Utilizing this concept of queuing theory, the collection of data patterns were generated. With the performance parameter of the system, the analysis of the data pattern had been made to diagnose the network anomaly. Performance analysis and results show the accuracy of the proposed scheme in detecting anomalies

——————————— ◆ ———————————

## 1 INTRODUCTION

Traffic anomalies such as failures and attacks are very common in today's computer networks. Identifying and then diagnosing and treating anomalies in a timely fashion are the fundamental part of day to day network operations. Without such kind of capability, networks are not able to operate efficiently or reliably. Exact identification and diagnosis of anomalies depends firstly on robust and timely data, and secondly on established methods for isolating anomalous signals within that particular data. As anomalies in communication networks are usually unavoidable, the early detection and classification of these faults is crucial to providing networking services with a higher level of availability and reliability. The main problem when trying to solve DOS attacks is detection of the attack which importance cannot be overstated. We need to detect an attack in progress as soon as possible for the various reasons. Firstly, the sooner the attack is detected prior to inflicting any damage, the more time would be taken by the system under the attack to implement some defense measures. Second, attack detection usually ascertains also the identity of those systems which participate in it. Such type of data is potentially useful for taking the legal action and prosecuting the guilty party. And finally, if the attack can be detected close to its own sources, corresponding filtering mechanism to that network can be turned on, dropping attack flows and moreover prevent the bandwidth waste. Of course, all these opportunities are available only if a given detection process is really doing what it is supposed to do, if it is really worthy. It has generally been hard to automate the anomaly identification process. A very vital step in improving the capability of identifying anomalies is to seperate and characterizes their important features. A road map for characterizing broad aspects of network traffic was outlined in [1]. In this paper, we restrict our focus to one aspect of the work and report results of a detailed analysis of network traffic anomaly. This analysis considers the time frequency characteristics of TCP flow, UDP flow collected at NS2 simulator having various nodes representing system nodes and having different traffic flows like File transfer protocol (FTP), Constant bit rate (CBR).

## 2 RELATED WORK

A serious type of network attack is Denial of Service, which is performed against the computer network to prevent legitimate users from accessing the services of compromised system. DOS attacks restrict a legal network user from performing his functions [2]. These attacks overcome the victim host to the point of unresponsiveness to the intended user of that host [3]. While, there are many options open to a malicious person having desire to launch a DOS attack, there are two basic classifications of attack; starvation of resources and bandwidth consumption [3], [4], [5], [6]. Resource starvation attacks [8] have their main focus on consuming all of the target resources so that they are unable to process any new requests for legitimate users. Examples are - Transmission Control Protocol (TCP) SYN flooding [3] consume all their victim's resources with half open requests for connection. Bandwidth consumption attacks are launched when an attacker sends large amount of data at the victim host than it is able to deal with, filling all communications channels with this data. For example, Internet Control Message Protocol (ICMP) flooding or User Datagram Protocol (UDP) flooding [3]. The situation is further complicated when it comes to Distributed Denial of Service

————————————————————

- *Neetu Singh is with Jaypee University Of Information Technology, Waknaghat, Solan, H.P, India.*
- *S.P.Ghrera, with Jaypee University Of Information Technology, Waknaghat, Solan, H.P, India.*
- *PranayChaudhuri is with Jaypee University Of Information Technology, Waknaghat, Solan, H.P, India.*

(DDoS) tools, like "trinoo" [7]. The [12] concerned about detecting DOS attacks using Support Vector Machines (SVMs). The key idea behind is to train SVMs using known discovered patterns (signatures) that represent DOS attacks. Using a benchmark data from a Knowledge Discovery and Data Mining (KDD) competition designed by DARPA (Defense Advanced Research Projects Agency), Anomaly detection systems use user profiles as the basis for detection; any variation from the normal user behavior is stated as intrusions [11], [12], [13], [14], [15], [16]. Several intrusion detection systems are built with human or without the human intervention. Ware and Steven Levy figured out the need for computer security [9], [10]. Identifying anomalies quickly and accurately is critical to the efficient operation of huge computer networks. Accurately characterizing the classes of anomalies greatly facilitates their identification; however, the subtleties and complexities of anomalous traffic can easily confound this process. In [17], an effective way of exposing anomalies is via the detection of a sharp increase in the local variance of the filtered out data. The traffic anomaly signals are evaluatedat different points within a network that is based on topological distance from the anomalous source or destination. .In [18], the efficiency of diagnosing network anomalies using concepts of statistical analysis and evidential reasoning had been investigated. Principle component analysis (PCA) is introduced in [19] and [20] to separate variations in traffic at any point into normal and anomalous components. In [21], Thottan et. al. abrupt changes are detected with the help of traffic management information base (MIB) counts obtained via the SNMP protocol using time series analysis. Second order statistics were used in [16] to investigate load anomalies in IP networks. Barford et al. [22] use pseudo-spline wavelets as the root to analyze the time limited to a small area normalized variance of the high frequency component to identify signal anomalies. Entropy-based methods [23], [24] identify anomalies by computing the entropy related with the probability distribution of the monitored traffic. Morover, analysis of the collected measurements in order to diagnose anomalous behaviour is still considered an unmet challenge and remains to be the focus of numerous ongoing research practice [25], [22], [19], [26], [21]. Bayesian networks are also applied for anomaly detection (e.g., by Ho et al. [28] and Auld et al. [27]).In this paper we diagnose the DOS attack or anomaly in the network by the means of performance parameter such as packet arrival, packet drop, bandwidth etc. and queuing theory.

## 3 OUTLINES OF PROPOSED ARCHITECTURE

In this section, we outlined the diagnosis approach employed in the proposed scheme for DOS attack at realtime. A schematic description of the proposed method typically implemented at a network node is presented in fig. 1, it is assumed here that the node is supported with the necessary performance monitoring and data measurements and collection tools. As shown, the scheme is divided into two main modules: an abrupt change detec-tion and an alarm/signal generator.

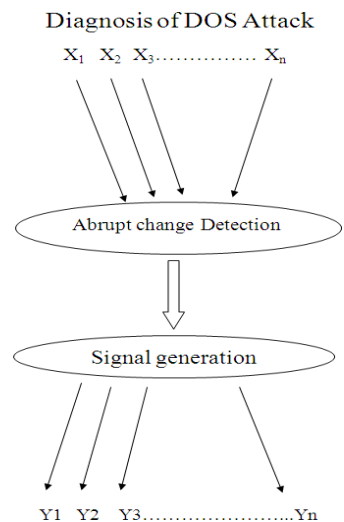

Fig. 1 Formal description of the proposed approach

The first module takes the performance parameter ($X_1$, $X_2$ $X_3$.............. $X_n$) of the system as the input and analyse the data patterns with respect to time and other performance variables. This module works on the concepts of queuing theory. Finally it passes the output to the signal generation module for further processing so that alarm ($Y_1$, $Y_2$ $Y_3$......................$Y_n$) could be generated when the attack is in progress.

## 4 ABRUPT DETECTION PHASE

A typical input to any network diagnosis system is a collection of continuous stream measurements of performance monitoring variables that measure the behavior of traffic flows (example, number of dropped packets and link utilization) as well as the nodes health (like, CPU utilization and memory load). This measurement can be represented as, for example pkt_ (packet number entering in the system) or be presented as set of MIB counts collected via SNMP. In order to perform the first step of abrupt change detection, we follow a exponential approach of measurements obtained from network monitoring variables. Network anomalies could be identified through the analysis of the spectral characteristics of the obtained data streams representing performance variables.

Queuing Theory: Every network must have some mechanism to store and forward the request arriving at the system. The memory space and allocation is also fixed. So it is certainly the key point for the attacker to launch the attack and somehow disable the server to provide the service to its legitimate users. Therefore queuing theory conceptually hold the power to detect some kind of DOS attacks like UDP flooding and TCP flooding attack. The simple architecture of queue is described in fig. 2.

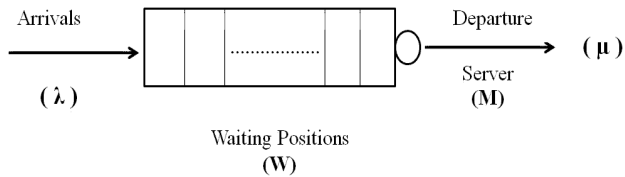

Fig. 2 Simple Queue architecture

It shows that λ represents the jobs arriving at the queue, having waiting time W till they got no response with server. The departure rate is μ, all are related with the single server only. Different types of queue models are defined to hold and manage the data. These are Poisson, exponential, priority, multiple servers etc. Each of these has its own basic mechanism to process the request on the basis of first come and first serve (most commonly used) , shortest job first and priority . As we are focusing on traffic analysis, the queue must support exponential data. And the process through which the requests are processed is first come first serve having single server and obviously finite buffer state.

A Single-Server Exponential Queuing System: Assume that requests arrive at a single-server in accordance with a Poisson process having rate λ That is, the time between successive arrivals are independent exponential random variables having mean 1/λ Each request, upon arrival, goes directly into service if the server is free and, if not, the request joins the queue. When the server finishes serving a node, the request from that node would leaves the system, and the request from next node in line, if there is any, enters service. The successive service times are supposed to be independent exponential random variables having mean 1/μ The preceding is called the M/M/1 queue. The two Ms refer to the fact that both the inter arrival and the service distributions are exponential (and thus memoryless, distributions are exponential (and thus memory less, or Markovian), and the 1 to the fact that there is a single server. Now in this paper performance parameter of the system is used to diagnose the DOS attack via using queue theory.

$$N = \lambda W \qquad (1)$$
$$Nq = \lambda Wq \qquad (2)$$

Result holds in general for virtually all types of queuing situations where

λ= Arrival rate of data that actually enter the system

N = Total number of data entered in system

W = Waiting period of the data

Nq = Waiting period of data in the particular queue

Data blocked and refuse entry into the system will not be counted in λ

Feature Analysis: The major goal of feature analysis is to select and extract those network features having the potential to discriminate the anomalous behaviors from normal network activities. Since most current intrusion detection systems use network flow data as their information sources, we focus on features in terms of flows. The following parameters were used to measure the entire network's behavior.

Performance parameter: Performance parameter are the parameters that analyze the system performance and some changes in them lead to variation in system performance.

The various parameters used in this approach are:-

 FlowCount. A flow consists of a group of packets/data going from a particular source to a particular destination over a time period. The network flow should at least include a source (having source IP, source port), a destination (having destination IP, destination port), IP protocol, number of bytes and number of packets. Flows could be considered as sessions between users and services. Since attacking behaviors are usually different from normal user activities, they may be detected by observing flow characteristics.

AveragePacketSize. The average number of bytes per packet is in a flow over a specific time period.

Parrivals_. The number of packet arrival at the particular queue between the two nodes. It is used for queue monitoring. The higher the number of packets arrived at the server , higher the chances of attack. The main focus of the attacker is to utilize the resources as much as possible.

Pdepartures_. The number of packet departures from the particular queue in between the two nodes is used for queue monitoring.

 Bandwidth_. The amount of bandwidth used by the system with respect to time. It is used to analyze the current bandwidth of the particular link. The bandwidth allocation is fixed. That is after a limit, the server is unable to respond to the requests.

Pdrops _.The number of packet drops between the nodes. Higher the number of packet drop and higher the amount of bandwidth utilization, Moreover the chances of attack such as flooding attack are higher.

# 5 SIGNAL PHASE

The abrupt detection phase captured the data patterns which were generated by queuing theory. After that it analyzed the performance parameter. In this context the main aim was to diagnose the anomalous behavior on the basis of these parameters. When the attack was undergoing, some signal had been sent to the administration immediately so that effective measures could be taken to stop the attack and to trace back the attackers.

FLOW BEHAVIOUR: To analyze the flow behavior, a sudden change or increase of the traffic characteristics had been checked on a certain threshold, whether it was an anomalous or a normal behavior towards the system. Hence alarm could be raised at that particular time, whenever the anomaly was detected in the system.

NUMBER OF PACKET DROPS: The resource depletion attack such as UDP, ICMP attack have their main focus on the usage of resource as much as they can. Which in turn make resources unavailable to its intended users. Now as the resources are fixed and limited. The attacker starts deprecating the resources. As we have concentrated on the UDP attack, here the attacker sends the large number of UDP packets to the server. This certainly effects the bandwidth and buffer utilization of the system. Due to this other requests would be blocked to reach the server for further processing. Hence, the attack had been launched for the time being. Now for that particular time period the bandwidth utilization should be maximum and the number of arrivals will be equal to the number of drops. Now we can say that number of arrivals is equal to number of drops under maximum bandwidth utilization and buffer overflow at the time of attack. In our experimental set up the maximum bandwidth utilization is up to 80-90%. If such kind of conditions were found to be true the signals or alarm was generated.

## 6  RESULT AND DISCUSSION

The experiment was carried out through network simulator 2 having various nodes, each acting as an independent computer in the network. Among all, one of the node acts as a server that serves the requests of all the other nodes. All nodes are flooded with traffic flow characteristics. In this paper, we have considered only TCP and UDP flows. Some nodes have CBR agents over UDP and other have FTP agent over TCP. M/M/1 queue theory was applicable to our research as there is a single server. The data patterns were generated through three UDP flooding attack, launched at different times. First attack was launched via port 21, second via port 5060 and third via port 1580. Then the second step was to measure the performance variable to diagnose the DOS attack. It is clear that at the time, when flooding attack was in progress the number of packet drops was increasingly very rapidly (fig. 3). This clearly shows that at the time of attack, the number of packet had been dropped are maximum.

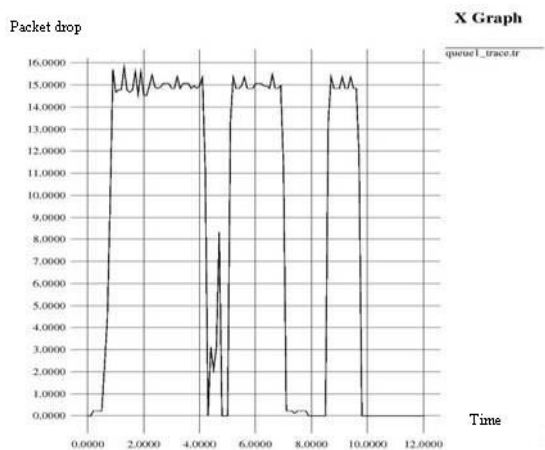

*The data generated on y axis is multiple of $10^3$

Fig. 3 Number of packet drop with respect to time

During the three attacks the bandwidth utilization was maximum. This was due to the fact that at the attack time, the number of requests was very high and the resources were fixed to serve the requests. Hence at this particular time, the bandwidth utilization was also maximum (fig. 4)

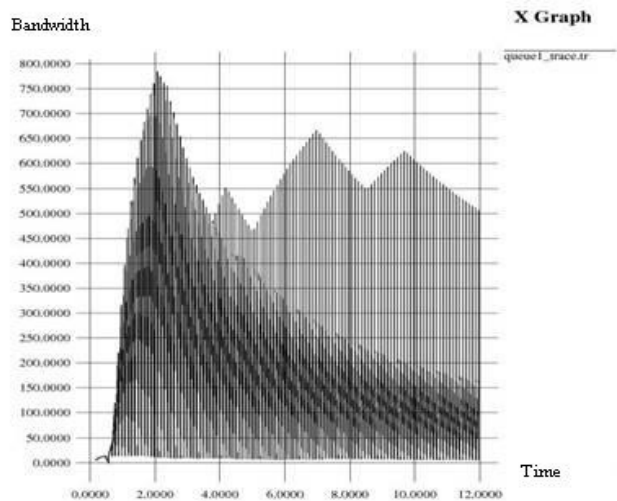

*The data generated on y axis is multiple of $10^3$

Fig. 4 Bandwidth utilization of the system with respect to time

Fig. 5 shows that at the time of attack the number of packet arrival was more and hence the bandwidth utilization too. As the number of packet arrival was more, more was the need for the bandwidth for further processing of these requests.

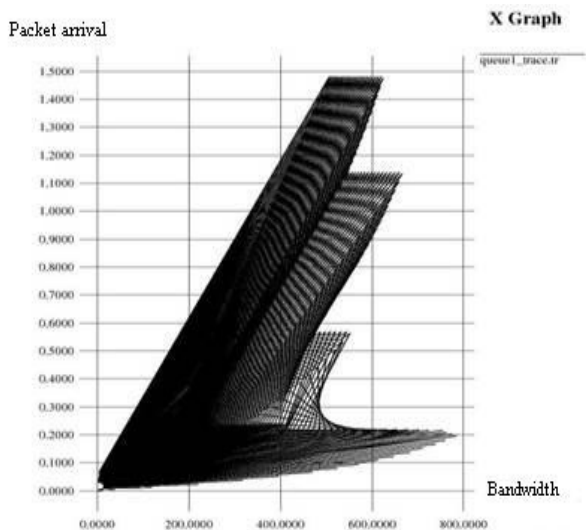

*The data generated on x and y axis is multiple of $10^3$

Fig. 5 Number of packet arrivals in the term of the bandwidth utilization

Fig. 6 shows that the packet drop was maximum when the bandwidth utilization was maximum. This is because at the time when the bandwidth utilization was maximum, the server was unable to process the requests and hence they had been dropped.

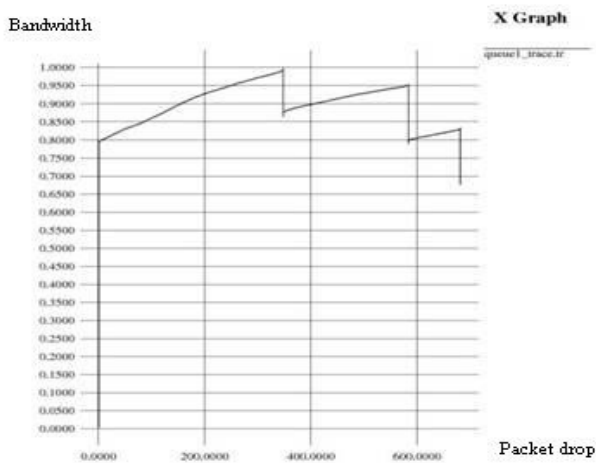

*The data generated on y axis is multiple of $10^6$

Fig. 6 Number of packet drops with respect to the bandwidth

The effect of the flooding attack on the FTP and CBR in the system having the conditions that resembles to the real world (Table 1 & 2).

| FLOOD ATTACK | Number of packet lost during attack (in %) |
|---|---|
| Flood 1 | 36.11% |
| Flood 2 | 35.26% |
| Flood 3 | 36.66% |

Table-1 Impact of UDP flood attack on CBR

| FLOOD ATTACK | Number of packet lost during attack (in %) |
|---|---|
| Flood 1 | 44.18% |
| Flood 2 | 46.12% |
| Flood 3 | 43.67% |

Table-2 Impact of UDP flood attack on FTP

After the evaluation of performance parameter, the signal generation was needed to diagnose the attack. Our focus was to diagnose the UDP flooding attack and the key target was the resources including bandwidth and the buffer only. That's why when the system reached its bandwidth utilization to a limit and number of packet arrivals is equal to the number of packet drops. We concluded it as

UDP flooding denial of service attack. At the time of attack, the flow of packets was maximum so that the buffer became overloaded .Hence; the requests from the intended users suffered and were forced to be dropped. And the alarms are generated to diagnose the attack when it was in progress The value of signal was varied only from 0 to1 (fig. 7). Zero means no attack and 1 means attack in progress. It is clear from this figure that attack was diagnosed successfully.

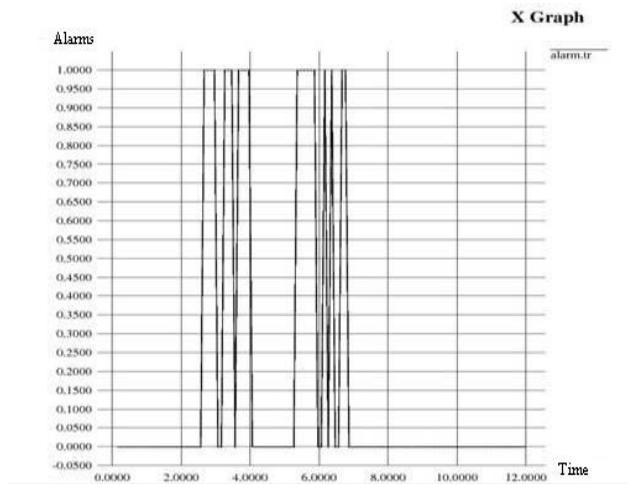

Fig. 7 Signal generation with respect to time

Finally, (table-3) shows the result and experiment inference of the experimental setup as a part of the research.

| Estimated parameter | % correct estimation |
|---|---|
| Operational output/result | 96% |
| Methodology | 95% |

Table-3 Result and method inference

# 7 CONCLUSION AND FUTURE WORK

The threat to organizations from network attacks is very common and real. Although, DOS attacks are seriously underrepresent in current research. There is a very real need for organizations to protect their technological and information resources from such type of attacks. Within this paper, we have presented the approach which focuses on detecting DOS beyond the conventional systems. This approach detects attacks on the communication medium that all traffic, whether valid or invalid, must traverse to reach the intended destination. We have presented a means by which detection of ongoing attacks can be achieved as soon as possible. To demonstrate the applicability of our approach, we have launched the UDP flooding DOS attack and diagnose the attack in progress by applying Queuing theory and then analyses the traffic flows via performance parameter. Our future work will concentrate on the issues such as applicability of this model to real world and on all types of DOS attack instead of flooding attack only.

**Neetu Singh** is M.Tech student of Jaypee University of Information Technology, Waknaghat, India. She is persuing her M.Tech with specialization in computer networks under the supervision of Satya Prakash Ghrera. She has published her research in leading international conference proceedings.

**Satya Prakash Ghrera** is currently an Associate Professor of Computer Science and Engineering at Jaypee University of Information Technology, Waknaghat, India. Prior to joining Jaypee University of Information Technology, he served the industry for 35 years. His research interests include design of Computer Networks, Computer and Network Security, Network Management, Microprocessors and Controllers, Integration of Computer Networks and Communication

Systems, Network programming, Management of Network based Real Time Information Systems, and Image Processing. He has published his research in leading international conference proceedings.

**Pranay Chaudhuri** is currently a Professor of Computer Science and Engineering at Jaypee University of Information Technology, Waknaghat, India. Prior to joining Jaypee University of Information Technology, Professor Chaudhuri was holding a Chair Professorship in Computer Science at the University of the West Indies, Cave Hill Campus, Barbados (2000-2009). Professor Chaudhuri has also held faculty positions at the Indian Institute of Technology at Kharagpur, James Cook University of North Queensland, University of New South Wales and Kuwait University. Professor Chaudhuri's research interests include Parallel and Distributed Computing, Grid Computing, Self-stabilization and Algorithmic Graph Theory. In these areas, he has extensively published in leading international journals and conference proceedings. He is also the author of a book titled, Parallel Algorithms: Design and Analysis (Prentice-Hall, Australia, 1992), and co-author of a book titled, Task Scheduling in Distributed Computing Environments (Lambert Academic Publishing, Germany, 2010). Professor Chaudhuri is the recipient of several international awards for his research contribution.

.